\documentclass[nofootinbib,floats,aps,superscriptaddress,preprint]{revtex4}
\usepackage{feynmf,slashed}

\newcommand{\beqa}{\begin{eqnarray}}
\newcommand{\eeqa}{\end{eqnarray}}
\newcommand{\beq}{\begin{equation}}
\newcommand{\eeq}{\end{equation}}

\newcommand{\lsim}{\lesssim}

\def\GeV{{\rm GeV}}
\def\MeV{{\rm MeV}}
\def\eV{{\rm eV}}
\def\tr{\mbox{Tr}}
\def\Tr{\mbox{Tr}}
\def\kon{k_{on}}

\newcommand{\Pl}{\rm Pl}

\def\addtech{Department of Physics,
Technion--Israel Institute of Technology,\\
Technion City, 32000 Haifa, Israel\vspace*{20pt}}
\def\addbu{Physics Department, Boston University, Boston,
MA 02215\vspace*{6pt}}
\def\addharvard{Jefferson Laboratory of Physics, Harvard University,
  Cambridge, MA 02138\vspace*{6pt}}

\begin{document}

\preprint{\vbox{\hbox{hep-ph/0506216}
\hbox{HUTP-05/A0031}\hbox{June, 2005}}}
\vspace*{1cm}

\title{Neutrino Constraints on Spontaneous Lorentz Violation}

\author{Yuval Grossman}\affiliation{\addharvard}\affiliation{\addbu}\affiliation{\addtech}
\author{Can Kilic}\affiliation{\addharvard}
\author{Jesse Thaler}\affiliation{\addharvard}
\author{Devin G. E. Walker}\affiliation{\addharvard}

\begin{abstract} \vspace*{8pt}
We study the effect of spontaneous Lorentz violation on neutrinos.  We
consider two kinds of effects: static effects, where the neutrino
acquires a Lorentz-violating dispersion relation, and dynamic effects,
which arise from the interactions of the neutrino with the Goldstone
boson of spontaneous Lorentz violation.  Static effects are well
detailed in the literature.  Here, special emphasis is given to the
novel dynamic effect of Goldstone-\v{C}erenkov radiation, where
neutrinos moving with respect to a preferred rest frame can
spontaneously emit Goldstone bosons.  We calculate the observable
consequences of this process and use them to derive experimental
bounds from SN1987A and the CMBR.  The bounds derived from dynamic
effects are complementary to --- and in many cases much stronger than
--- those obtained from static effects.
\end{abstract}

\maketitle

\section{Introduction}
Neutrinos provide an interesting laboratory for studying possible
violations of Lorentz invariance. First, neutrino oscillation
experiments indicate that neutrino mass differences are much smaller
than $1\;\eV$.  Thus, neutrinos should be sensitive to small
deviations from relativistic energy-momentum relations.  Second,
because neutrinos are so weakly interacting, they are sensitive to
physics over very long times and distances.  Neutrinos from distant
astrophysical sources like supernovae can probe very small effects
that accumulate throughout the time of travel.  Finally, neutrinos
constitute a significant amount of the energy of the universe during
the time the Cosmic Microwave Background Radiation (CMBR) was
generated.  During this epoch, neutrinos are decoupled from the
baryon-photon plasma and free-stream from over- to under-dense
regions.  Any deviation from this picture can affect the CMBR signals
we observe today.

While the specific origin of Lorentz violation (if any) is unknown,
there are some model-independent statements we can
make.\footnote{There is a growing literature on specific
Lorentz-violating modifications of gravity.  See, for example
\cite{Arkani-Hamed:2003uy,Rubakov:2004eb,Gripaios:2004ms,Dubovsky:2004sg,Eling:2004dk,Bluhm:2004ep,Graesser:2005bg,Libanov:2005vu}.}
If general relativity is the correct description of gravity up to the
Planck scale, then Lorentz violation must be spontaneous.  The reason
is that any operator that breaks Lorentz invariance necessarily
violates space-time diffeomorphisms, which is the gauge symmetry of
gravity.  As with all gauge symmetries, diffeomorphisms (and hence
Lorentz invariance) can only be broken spontaneously.  A spontaneously
broken symmetry implies the existence of a Goldstone boson, though in
a Lorentz-invariant context, this Goldstone boson would be ``eaten''
by the gauge field and all physical polarizations would become
massive.  However, if we break both diffeomorphisms and Lorentz
invariance, the physical Goldstone boson can be exactly massless and
lead to novel interactions in the infrared.  At minimum, the Goldstone
boson of spontaneous Lorentz violation will mix with gravity, which
may or may not lead to measurable modifications of gravitational
physics.

In this paper, we focus on possible direct couplings between neutrinos
and the Lorentz-violating sector which survive in the $M_{\Pl}
\rightarrow \infty$ limit.  In order to do explicit calculations, we
have to make some assumptions about the Lorentz-violating sector.  We
consider two models where a scalar field acquires a Lorentz-violating
vacuum expectation value (vev), ``ghost condensation''
\cite{Arkani-Hamed:2003uy} and ``gauged ghost condensation''
\cite{Gauge-ghost}.  In addition to providing a systematic way for
studying Lorentz violating effects, ghost condensation yields a
consistent infrared modification of gravity
\cite{Arkani-Hamed:2003uy,Mukohyama:2005rw} and has been used as a
novel model of inflation
\cite{Arkani-Hamed:2003uz,Senatore:2004rj}.  Direct
couplings between the ghost condensate and the standard model were
considered in \cite{Arkani-Hamed:2004ar}, where several dynamic
phenomena involving the Goldstone boson were identified and studied.
We extend this work by investigating how \v{C}erenkov radiation from
neutrinos into the Goldstone field can be used to probe spontaneous
Lorentz violation.

In general, couplings between any Lorentz-violating sector and
neutrinos produce two kinds of effects which we categorize as
``static'' and ``dynamic.''  The static effects arise from
Lorentz-violating vevs, which define a preferred ``ether'' frame. We
choose this frame to be the CMBR rest frame since in the context of
ghost condensation, the two frames are aligned due to Hubble
friction \cite{Arkani-Hamed:2003uy}.  The dominant static effect is a
modification of the neutrino dispersion relation, which may or may not
be CPT-violating.  Lorentz-violating static effects in the neutrino
sector were studied, for example, in
\cite{Coleman:1998ti,Kostelecky:2003cr,Kostelecky:2003cr-b,Kostelecky:2004hg,Glashow:2004im,Hooper:2005jp},
and there is a large literature on the effect of preferred frames on
standard model fields
\cite{Colladay:1998fq,Colladay:1996iz,Coleman:1998ti,LorentzOne,LorentzTwo,Bluhm:2003ne}.

Dynamic effects arise from couplings between neutrinos and the
Goldstone boson of spontaneous Lorentz violation.  In particular,
neutrinos can lose energy while traveling in vacuum as they emit
Goldstone-\v{C}erenkov radiation.  The parameters that control the
interactions with the Goldstone boson are related to those that enter
into static effects, thereby providing complementary tests of
Lorentz invariance.  We emphasize that while the details of the
dynamic effects depend on the specific interactions of the Goldstone
boson, the existence of a Goldstone boson is robust, so we expect our
results to have analogs in any theory where neutrinos are directly
coupled to a Lorentz-violating sector.

Spontaneous Lorentz violation is very similar to the well known matter
effect in neutrino oscillations \cite{nu-rev}. In both cases the
neutrino travels in a non-trivial background that breaks Lorentz
invariance.  For the matter case, the effect is generated by the weak
interaction between the neutrinos and the medium they travel through,
whereas in ghost condensation, the effect exists in vacuum.  In both
cases, the Lorentz-violating background generates a static effect,
namely an effective ``mass'' for the neutrino.  There is, however, a
fundamental difference. For the matter effect the full theory of the
weak interaction and the values of its parameters are known.  In
particular, inelastic interactions between neutrinos and matter
(\emph{i.e.}\ phonon production) are usually negligible. In ghost
condensation, on the other hand, dynamic interactions with the
Goldstone boson are important for generic values of the parameters in
the effective theory.  Also, in the case of matter effects, one
usually considers only the lowest dimension operators since higher
dimension ones are known to be much smaller.  In ghost condensation,
however, one needs to consider higher dimension operators as well, as
the lowest dimension operators may be forbidden (or suppressed) by
symmetries of the underlying ultraviolet theory.

By assuming that the Lorentz-violating sector is (gauged) ghost
condensation, we are focusing only on the case where Lorentz symmetry
is spontaneously broken to rotational symmetry.  In the most general
case, rotational symmetry can also be broken, yielding additional
static effects and presumably new Goldstone bosons.  As far as
neutrinos are concerned, violations of rotational invariance generate
helicity flip tensor operators, which is equivalent to the effect of
neutrino traveling in a magnetic field \cite{book}. The implication of
such interactions in matter were studied in
\cite{Bergmann:1999rz}. For the Lorentz-violating case, such couplings
were discussed in \cite{Kostelecky:2003cr,Kostelecky:2003cr-b}. For
Dirac neutrinos, these Lorentz-violating tensor operators conserve
total lepton number but generate left-right oscillations, such that a
left-handed active neutrino will oscillate into a sterile right-handed
neutrino. For Majorana neutrinos, they violate lepton number and must
be off-diagonal in flavor space, generating oscillations between
neutrinos and anti-neutrinos of different flavors. For the remainder
of the paper, we do not consider tensor interactions any further.

In the next section, we first summarize the formalism of (gauged)
ghost condensation and show how the ghost condensate couples to
neutrinos.  We then estimate the sizes of various Lorentz-violating
static and dynamic effects on neutrinos, using SN1987A and the CMBR to
constrain Goldstone-\v{C}erenkov radiation.  We find that the two
effects yield complementary bounds on the size of Lorentz-violating
operators.  The main results are listed in the body of the paper while
the details of our calculations are given in the appendices.

\section{Formalism}

\subsection{Ghost Condensation}
\label{gcformal}

The basic ingredient of ghost condensation is a scalar field $\phi$
which acquires a Lorentz-violating vev.\footnote{In a gravitational
context, the low energy physics of ghost condensation is uniquely
described by the spontaneous breakdown of space-time diffeomorphisms
to spatial diffeomorphisms, whether or not there actually exists a
field $\phi$ that accomplishing this breaking pattern
\cite{Arkani-Hamed:2003uy,Arkani-Hamed:2004ar}.}  The field $\phi$ has
a wrong sign kinetic term near the origin, but this ghost-like
instability is stabilized at a non-zero value of $(\partial\phi)^2$.
There is a global shift symmetry acting on $\phi$ as $\phi \rightarrow
\phi + c$, so the leading terms in its effective Lagrangian are
\beq  \label{gclagrange}
\mathcal{L} = \frac{1}{8M^4}\left((\partial_\mu \phi)^2 -
M^4 \right)^2 - \frac{\beta}{2M^2}(\partial^\mu \partial_\mu \phi)^2 +
\ldots, 
\eeq
where $M$ defines the characteristic scale of Lorentz violation and
naturalness suggests $\beta \sim 1$.  Going to a preferred ``ether''
rest frame, we assume that $\partial_\mu \phi$ acquires a vev in the time
direction
\beq
\langle \partial_\mu \phi \rangle = \delta_\mu^0 M^2,
\eeq
such that Lorentz invariance is broken down to rotational invariance.
It is convenient to write
\beq \label{vevphi}
\phi \equiv M^2 t + \pi,
\eeq
where $\pi$ is the physical Goldstone boson of the broken symmetry.
Expanding to leading order in the Goldstone energy $E_\pi$ and momenta
$k$, the Goldstone has the following Lorentz-violating dispersion
relation
\beq \label{pion-dis}
E_\pi^2=\beta \frac{k^4}{M^2} + O(k^6/M^4).
\eeq
This novel dispersion relation leads to interesting dynamics between
neutrinos and the Goldstone.  Note that the effective theory for $\pi$ is
applicable only for $k \ll M$, so an unspecified UV completion is
necessary to accurately handle large Goldstone energies.  The bound on $M$
from mixing between $\pi$ and the Newtonian potential were studied in
\cite{Arkani-Hamed:2003uy}, where it was found that
\beq
M \lsim 10\; \MeV.
\eeq
Because neutrinos often have energies in excess of $10$ $\MeV$, bounds
on ghost condensation from energetic neutrinos should be regarded as
$\mathcal{O}(1)$ estimates.

\subsection{Gauged Ghost Condensation}

A simple modification to ghost condensation is to gauge the $\phi
\rightarrow \phi + c$ shift symmetry.  The resulting theory of gauged
ghost condensation has the nice feature that the scale of spontaneous
Lorentz violation $M$ can be taken to be much higher, even well above
the electroweak scale, without violating experimental bounds on
gravity \cite{Gauge-ghost}.  Let the gauge transformation act on
$\phi$ and the new $U(1)$ gauge field $A_\mu$ as
\beq
\phi \rightarrow \phi + M \alpha, \qquad A_\mu \rightarrow A_\mu - \partial_\mu \alpha,
\eeq
such that the gauge invariant derivative is
\beq
D_\mu \phi = \partial_\mu \phi + M A_\mu.
\eeq
As in ghost condensation case, we choose $D_\mu \phi$ to acquire a vev
in the time direction.  Minimally coupling the gauge field to equation
(\ref{gclagrange}), the leading effective Lagrangian for $\phi$ and
$A_\mu$ is
\beq  \label{ggclagrange}
\mathcal{L} = -\frac{1}{4g^2}F_{\mu\nu}^2 + 
\frac{1}{8M^4}\left((D_\mu \phi)^2 - M^4 \right)^2 - 
\frac{\beta}{2M^2}(\partial^\mu D_\mu \phi)^2 - 
\frac{\beta'}{2M^2}(\partial_\mu D_\nu \phi)^2  + \ldots,
\eeq
where $g$ is the gauge coupling, and we expect $\beta, \beta' \sim
1$. (The $\beta$ and $\beta'$ terms appear since there is an ambiguity
in covariantizing the $\beta$ term in (\ref{gclagrange}).) There is a
convenient ``unitary'' gauge where $\phi \equiv M^2 t$,
\beq
D_\mu \phi \equiv M^2 \delta_\mu^0 + M A_\mu.
\eeq
In this gauge, the field $A_0$ has mass $gM$ and can be integrated
out, and we can absorb $\beta'$ into a redefinition of $g$ and
$\beta$.  The remaining three degrees of freedom in $A_i$ are
comprised of two transverse and one longitudinal mode.  To leading
order in $k/M$ and assuming $g \ll 1$, the transverse modes have the
Lorentz-invariant dispersion relation $E^2 = k^2$ and are irrelevant
to our discussion.  The longitudinal mode, however, has the following
Lorentz-violating dispersion relation:
\beq
E_L^2 = \beta \left(g^2 k^2 + (1-\beta g^2)^2 \frac{k^4}{M^2} \right) + \mathcal{O}(k^6 /M^4).\label{gauge-disp}
\eeq
In order to safely integrate out $A_0$, we have to be in the regime $k
\ll gM$.  In this limit, the normalized polarization vector and
dispersion relation are
\beq
\epsilon_i^L(k) = g k_i / |k|, \qquad  E_L^2 = \beta g^2 k^2,\label{long-polarization}
\eeq
and $\sqrt{\beta} g$ is essentially the phase velocity of the
Goldstone mode.  Note that in order to achieve large values of $M$
without drastically modifying gravity in the infrared, $g \gg
M/M_{\Pl}$ \cite{Gauge-ghost}, which is easily satisfied for any $M$
below the GUT scale.  Details of the construction of gauged ghost
condensation appear in appendix \ref{sec:ggc}.

\subsection{Couplings to Neutrinos}
\label{couplingneutrino}

We now consider direct couplings between neutrinos and the
Lorentz-violating sector.  In the following discussion, we neglect all
other standard model fermions, making the implicit assumption that
Lorentz-violating couplings to left-handed electrons and neutrinos are
not necessarily related by electroweak symmetry.  The lowest dimension
current involving just left-handed neutrinos is
\beq
J_\mu  = \bar\nu_L\gamma^\mu \nu_L,
\eeq
so the leading interaction with the ghost condensate is the dimension
five coupling
\beq \label{lowest}
\mathcal{L}_{\rm int} = \frac{1}{F}J^\mu \partial_\mu \phi,
\eeq
where $F$ sets the mass scale for interaction between neutrinos and
the Lorentz-violating sector.  Note that if the neutrino were exactly
massless, this interaction could be removed by a field redefinition on
$\nu_L$ because left-handed neutrino number would be conserved.
Therefore, any process involving equation (\ref{lowest}) must be
proportional to the neutrino mass, either a Dirac mass or a Majorana
mass.  Expanding $\phi$ around its vev as in equation (\ref{vevphi}),
we have two terms that yield complementary Lorentz-violating effects:
\beq \label{lowest-int}
{\cal L}_{\rm int}={M^2 \over F} J^0 + {1 \over F} J^\mu \partial_\mu
\pi.
\eeq
The first term gives rise to well-known static effects that change the
dispersion relation of the neutrino.  The second term generates
dynamic effects between the neutrino and the Goldstone boson which are
the focus of this paper.  Note that the same coupling $F$ sets the
size of the static and dynamic effects, and this is generically true
in any theory of spontaneous Lorentz violation
\cite{Arkani-Hamed:2004ar}.  Because Lorentz-invariance is violated,
there is no {\it a priori} reason that $J^0 \partial_0 \pi$ and $J^i
\partial_i \pi$ should share the same coupling constant.  In the
context of ghost condensation, however, the couplings are split
by the dimension nine operator
\beq
\mathcal{L}_{\rm int} = 
\frac{C_9}{F^5} J^\mu \partial_\mu \phi \partial_\nu \phi \partial^\nu \phi,
\eeq
where naturalness suggests $C_9 \sim 1$, and therefore we generically
expect the splitting to be suppressed. We explicitly checked that for
$C_9=1$ we can safely neglect any difference between the couplings to
the time and spatial components of $J^\mu$.

Next, consider coupling $J^\mu$ to the gauged ghost condensate:
\beq \label{gaugedint}
\mathcal{L}_{\rm int} = \frac{1}{F} J^\mu D_\mu 
\phi \quad \Rightarrow \quad \frac{M^2}{F} J^0 + \frac{M}{F} J^i A_i,
\eeq
where we have gone to $\phi \equiv M^2 t$ gauge and integrated out
$A_0$ as in the previous section.  If the neutrino is exactly
massless, then the static effect generated by the $J^0$ term can be
rotated away by a $\nu_L$ field redefinition.  However, even if we
isolate the $A_i$ longitudinal mode by setting $A_i \sim \partial_i
\pi$, the dynamic coupling $J^i A_i$ cannot be removed by a field
redefinition.  We see that in gauged ghost condensation, dimension
five Lorentz violation survives in the $m \rightarrow 0$ limit where
$m$ denotes the mass of the neutrino.

Note that the couplings in equations (\ref{lowest}) and
(\ref{gaugedint}) can be forbidden by a $\phi \rightarrow -\phi$ (and
$A_\mu \to -A_\mu$) symmetry in the fundamental theory above the scale
$M$, so we consider the next most relevant set of operators that would
give rise to Lorentz violation even if the dimension five couplings
vanished.  Furthermore, we consider only operators that survive in the
massless neutrino limit. Up to total divergences and terms that can be
removed by field redefinitions in the massless limit, the leading
operators are the following dimension eight operators
\begin{eqnarray} \label{dim-eight}
&&\mathcal{L}_1=\frac{1}{F_1^{4}}T^{\mu\nu}
(\partial_{\mu}\phi)(\partial_{\nu}\phi),\\
&&\mathcal{L}_2=\frac{1}{F_2^{4}}T_\mu^\mu
(\partial_\nu\phi)(\partial^\nu\phi),\nonumber \\
&&\mathcal{L}_3=\frac{1}{2F_3^{4}}
J^\mu(\partial_{\mu}\phi)(\partial_\nu\partial^\nu\phi), \nonumber
\end{eqnarray}
where the energy-momentum tensor $T^{\mu\nu}$ for a massless fermion is
defined as usual
\beq
T^{\mu\nu}={1 \over 2} \bar \nu_L (\gamma^\mu
\partial^\nu + \gamma^\nu
\partial^\mu) \nu_L -  \eta^{\mu\nu} \bar\nu_L \gamma^\rho
\partial_\rho\nu_L.
\eeq
Note that unlike the dimension five operator, there is no symmetry
that can forbid these dimension eight operators. In terms of the
Goldstone, we obtain the following new interactions from
$\mathcal{L}_1$
\beq \label{dim-eight-int}
\mathcal{L}_1=\frac{M^4}{F_1^4} T^{00} + 2 \frac{M^2}{F_1^4}
T^{0\mu} \partial_{\mu}\pi,
\eeq
where as before, the first term gives a static effect and the second
one dynamic couplings to the Goldstone.  There are no bounds on $F_2$
and $F_3$ from static effects because when we set $\phi$ to its vev,
${\cal L}_3$ vanishes and ${\cal L}_2$ simply renormalizes the
neutrino kinetic term.  The dynamic couplings are
\beq \label{dim-eight-int-23}
\mathcal{L}_2= \frac{2M^{2}}{F_2^{4}}
 T \partial_{0}\pi,
\qquad
\mathcal{L}_3=\frac{M^{2}}{2 F_3^{4}}
 J^0 \partial^{\mu} \partial_\mu \pi.
\eeq

In the case of gauged ghost condensation, we can covariantize
(\ref{dim-eight}) by replacing $\partial_\mu \phi$ with $D_\mu
\phi$. This is a straightforward procedure and we do not give the
details here.

\section{Review of Static Effects}

We have seen how the vacuum expectation value for the ghost condensate
gives rise to the static interactions in equations (\ref{lowest-int})
and (\ref{dim-eight-int}).  In this section, we review how these terms
generate Lorentz-violating dispersion relations for neutrinos and
quote the strongest known bounds on sizes of these operators.  Because
these interactions may not be flavor-universal, we add flavor indices
to the currents $J^{0}$ and $T^{00}$.  The static Lorentz-violating
Lagrangian is ($i$ and $j$ are flavor indices, not spatial indices)
\beq
\label{statictotal}
\mathcal{L} = \bar{\mu}^{ij} J_{ij}^0 + \bar{a}^{ij} T_{ij}^{00}, 
\qquad \bar{\mu}  
\equiv {M^2 \over F}, \qquad \bar{a} \equiv \frac{M^4}{F_1^4}.
\eeq
There is a subtlety in the definition of $J^0$ and $T^{00}$,
especially in the presence of a Dirac mass.  In the previous section,
we defined $J^0$ and $T^{00}$ to include left-handed projection
operators, but there is no reason why right-handed neutrinos would not
also feel the effect of Lorentz violation.  However, to leading order
in the neutrino mass matrix $\bar{m}^2$ and the Lorentz-violating
parameters $\bar{\mu}$ and $\bar{a}$, we can ignore additional
Lorentz-violating coupling to right-handed neutrinos.  The effective
Hamiltonian for left-handed neutrinos in the presence of equation
(\ref{statictotal}) is
\beq \label{effH}
H_{ij}=\left(|p| \pm \mu+ a |p|+ {m^2 \over 2 |p|}\right)
\delta_{ij} \pm \mu_{ij} + a_{ij} |p| + {m^2_{ij} \over 2 |p|} + \cdots.
\eeq
where the $+\mu$ is for left-handed neutrinos and the $-\mu$ is for
right-handed anti-neutrinos.  Here we use the notation that for a
general $n\times n$ matrix $\bar X$,
\beq
X={1 \over n}\tr\,\bar X_{ij}, \qquad
X_{ij}= \bar X_{ij}-X\, \delta_{ij}.
\eeq
Of course, when neutrinos travel in matter, one should also include the
standard matter effect in (\ref{effH}) \cite{nu-rev}.

The flavor-universal terms $m^2$, $\mu$, and $a$ affect neutrino
kinematics, while the non-universal terms $m^2_{ij}$, $\mu_{ij}$, and
$a_{ij}$ dominantly affect neutrino oscillations. The effects from
Lorentz-invariant neutrino masses are well studied and we will not
elaborate on them further. In the following, we study flavor-universal
and flavor-non-universal Lorentz-violating static effects.

\subsection{Universal Terms}

The universal Lorentz-violating terms affect the kinematics of the
neutrinos, and one effect is to shift their group velocity away from
the speed of light. For simplicity, we set the neutrino mass to zero
and from (\ref{effH}) we find
\beq
v_g = 1+a.
\eeq
We confirm the known result that $\mu$ does not affect the group
velocity. We see that the group velocity depends on $a$ and is energy
independent. While we do not consider them here, higher dimensional
Lorentz-violating operators can yield energy dependent group
velocities.

We can place bounds on $a$ by considering experimental situations
where the neutrino velocity is well measure over long time scales.  At
present, the best bounds come from supernova neutrinos. The only
supernova which has been observed in neutrinos is SN1987A
\cite{book}. This supernova was at a distance of about $1.7 \times
10^5$ light years, which corresponds to travel time for the neutrinos
of about
\beq
t_{87A}\sim 5 \times 10^{12}\;{\rm sec} \sim 10^{37}{\rm~GeV}^{-1}.
\eeq
The detected neutrino energies, of the order of a few tens of MeV,
were in the expected range.  The best bound we can get on the
universal shift in the neutrino velocity can be derived by comparing
the time of the neutrino signal to the light signal for SN1987A.  The
data indicates that the neutrino arrived within one day of the light,
and therefore we have the bound \cite{Stodolsky:1987vd}
\beq  \label{staticbound}
|v_g-1| \lsim 10^{-8} \qquad \Rightarrow \qquad
|a| \lsim 10^{-8}.
\eeq

Another consequence of modifying neutrino dispersion relations is that
decay kinematics are modified. That is, the phase space factor for
decays with neutrinos in the final state will be modified. In general,
this can affect both the decay rates and the way these rates scale
with the energy of the decaying particle. Such considerations were
recently used in \cite{DiGrezia:2005qf} in order to derive bounds from
beta and pion decays. The former bound is stronger and gives a bound
on the Lorentz-violating parameter $\mu$:
\beq \label{beta-bound}
\mu \lsim 10^{-6} \;\GeV.
\eeq
Note that \cite{DiGrezia:2005qf} did not quote a bound on $a$ from
these consideration. Bounds on $\mu$ and $a$ can also be derived from
from comparing lifetime measurements at different energies.  Since the
Lorentz-violating term proportional to $\mu$ violates CPT, bounds on
$\mu$ can be obtained by comparing lifetime measurements between
particles that decay to neutrinos and anti-particles that decay to
anti-neutrinos. While a detailed study of such bounds is needed, we
do not expect the bounds to be much stronger than those we presented.

\subsection{Non-universal Terms}

Much stronger bounds can be derived on Lorentz-violating couplings
that violate flavor because these couplings affect oscillation
probabilities for neutrinos.  Thus, solar, atmospheric and terrestrial
neutrino experiments are all sensitive to these couplings.

Rough estimates of the bounds on Lorentz-violating couplings can be
extracted by looking at the sensitivity for a given experiment to
neutrino mass effects.  That is, using equation (\ref{effH}), we can
use the known sensitivity to $\Delta m^2/E$ as an estimate to the
sensitivity to $\mu$ and $a E$. From solar neutrino experiments and
Kamland, which detect neutrinos with energy of order few $\MeV$ and
are sensitive to $\Delta m^2
\sim 10^{-4}\;\eV^2$ we estimate for $i \ne e$
\beq \label{solarbound}
\mu_{ei} \lsim {\Delta m^2 \over E} \sim 10^{-20} \;\GeV,
\qquad
a_{ei} \lsim {\Delta m^2 \over E^2} \sim 10^{-18}.
\eeq
For atmospheric neutrinos a detailed study was done in
\cite{Gonzalez-Garcia:2004wg,Battistoni:2005gy}. 
In a two generation approximation, it was found
\beq \label{ANbound}
\mu_{\mu\tau} \lsim 10^{-22} \;\GeV,
\qquad
a_{\mu\tau} \lsim  10^{-24}.
\eeq

We finally mention that bounds on non-universal Lorentz-violating
couplings, $\mu_{ii}-\mu_{jj}$ and $a_{ii}-a_{jj}$ (for $i \ne j$) can
be derived in a similar way. The numerical values of the bounds are
the same order as magnitude as those derived above on off-diagonal
couplings.

\section{Dynamic effects}
In addition to static effects arising from the vev of the ghost
condensate, we can derive complementary bounds on spontaneous Lorentz
violation by studying fluctuations around the vev, namely the $\pi$
field.  As shown in section \ref{gcformal}, $\pi$ is a massless scalar
field,\footnote{A definition of massless is ambiguous in a
Lorentz-violating theory. Here we use the term ``massless field'' to
denote a field that has gapless excitations.} and if it is coupled
directly to the standard model we would expect to get constraints from
fifth-force measurements as in the case of
axions~\cite{Moody:1984ba,Duffy:2005ab}. In this paper, however, we
restrict ourselves to couplings of the Goldstone to neutrinos, and
because there is no way (yet) to assemble large coherent neutrino
sources, we will rely on the novel effect of ether \v{C}erenkov
radiation to derive bounds on neutrino-Goldstone couplings.  Further
bounds from astrophysical considerations will be considered in
\cite{Next-paper}. As far as the neutrino-Goldstone interactions are
concerned, bounds from star cooling arguments are much weaker than
those we report here.

Neutrinos are relativistic and interact weakly with matter, so any
interesting bounds on their coupling to the Goldstone field is likely
to arise from processes that involve large distances and times. We
consider two such dynamical effects in detail in the context of
SN1987A and the CMBR.  Using (\ref{pion-dis}), the phase velocity of
the Goldstone in the preferred rest frame of the universe is
\beq v_\pi
= \sqrt{\beta} \frac{k}{M}.
\eeq
Therefore, a neutrino with any non-zero momenta will always be
traveling faster than most $k$ modes of the Goldstone field, making
\v{C}erenkov radiation from neutrinos into the Goldstone kinematically
possible. As we will show, this has two important consequences.
First, the emission of a single Goldstone quantum will deflect a
relativistic particle by a large angle, enough to completely change
its original path of travel. Second, the particle will lose energy,
which is a cumulative effect over several emissions. We use the first
fact to study SN1987A where we have a handle on the number of
neutrinos involved in a process, while we use the second fact to study
the CMBR where we have knowledge about the energy stored in
cosmological neutrinos.

The rate of energy loss due to Goldstone radiation from electrons was
calculated in \cite{Arkani-Hamed:2004ar}, where it was estimated
that no observable effects could be seen from ether \v{C}erenkov
radiation. Our case is quite different in that the calculation in
\cite{Arkani-Hamed:2004ar} is non-relativistic and classical,
whereas we do the calculation relativistically and in the quantum
theory.  Also, while the emission rate and the rate of energy loss are
astronomically small, neutrinos compensate by traveling astronomical
distances on astronomical time-scales.

\begin{figure}[t]
\begin{tabular}{c}
\begin{fmffile}{decay}
\begin{fmfgraph*}(80,60)
     \fmfleft{in1}
     \fmfright{out1,out2}
     \fmflabel{$\nu(p)$}{in1}
     \fmflabel{$\nu(q)$}{out1}
     \fmflabel{$\pi(k)$}{out2}
     \fmf{fermion}{in1,v}
     \fmf{fermion}{v,out1}
     \fmf{ghost}{v,out2}
\end{fmfgraph*}
\end{fmffile}
\end{tabular}
\vspace*{11pt}
\caption{The Feynman diagram for Goldstone-\v{C}erenkov
radiation from neutrinos. \label{decayfig}}
\end{figure}
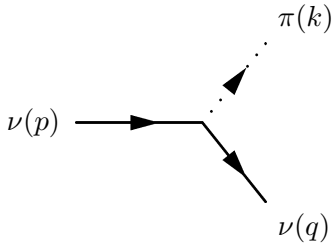

The \v{C}erenkov emission process we consider is $\nu \to \nu \pi$ as
in figure \ref{decayfig}. We take the initial and final neutrinos to
be the same species, and thereby do not consider flavor changing
vertices. This is justified as the experimental bounds on neutrino
mixing are much stronger than the kind of bounds we can derive from
our considerations. Furthermore, the main point in our analysis is
that when a neutrino emits a Goldstone it is deflected from its
original path of travel; whether it also changes flavor does not
affect our results. We use (\ref{lowest-int}) and
(\ref{dim-eight-int}) as the starting point of our calculation, the
details of which we present in appendix \ref{sec:calcappen}. Here we
quote only the final results relevant for putting bounds on sizes of
these operators.  Neutrinos are in the ultra-relativistic kinematic
regime and we only keep only the leading terms in an expansion
in powers of the neutrino mass.

The quantities that are of greatest interest for experimental bounds
are the emission rate $\Gamma$ --- the inverse of the average time for
a neutrino of a given momentum $p$ to emit a Goldstone quantum --- and
the rate of energy loss $-dE/dt$. We also list $\langle
\cos\varphi\rangle$, which is the average
deflection angle for a single emission event in order to show that it
is large enough to completely change the trajectory of the emitting
particle. All the results we quote should be taken as the leading
order results in $E/M$ where $M$ is the scale of spontaneous Lorentz
violation and $E$ is the energy of the initial neutrino which does not
change appreciably during a single emission.  For the case of gauged
ghost condensation we also work to leading order in the gauge coupling
constant $g$.  As already mentioned, all effects proportional to the
neutrino mass are already suppressed in the relativistic regime, so we
drop any subleading mass effects.  Finally, for the purposes of this
section, we can neglect any changes to the neutrino dispersion
relation from the static effects in the previous section, as $\mu/E$
and $a$ are smaller than any other quantity appearing in our
calculation.

For simplicity, we quote the results separately in different
scenarios, which is justified since the operators do not interfere.
We also substitute factors of $F$ in favor of $\mu$ and $a$ to simplify the
comparison of static and dynamic effects.  We begin by coupling the
ghost condensate with the dimension-five operator of equation
(\ref{lowest-int}) for which we find
\beq \label{rate-5}
\Gamma={1\over  4 \pi \sqrt{\beta}} {m^2\over M^3}\,\mu^2, \qquad
-{d E \over dt} = {1\over 2\pi}{m^2 E^2 \over M^4}\,\mu^2 , \qquad
\langle \cos\varphi\rangle = \frac{4E}{3M}\sqrt{\beta}\sim 0.
\eeq
Note that these rates are suppressed by the neutrino mass.  For this
reason we also consider the effects of the dimension eight operators
in equations (\ref{dim-eight-int}), which, even though they are
suppressed by higher powers of a high scale $F$, are not mass
suppressed. We find
\beq \label{rate-8}
\Gamma=\frac{E^{4}}{12\pi\sqrt{\beta} M^{3}}\,a^{2}, \qquad
-\frac{dE}{dt}=\frac{E^{6}}{6\pi M^{4}}\,a^{2},\qquad \langle
\cos\varphi\rangle=\frac{8E}{5M} \sqrt{\beta} \sim 0.
\eeq
For the operators of (\ref{dim-eight-int-23}) we find that ${\cal
L}_2$ does not contribute, because it is proportional to the trace of
the energy-momentum tensor which vanishes for massless on-shell
neutrinos.  For ${\cal L}_3$, due to our definition of $F_3$, the
results can be obtained from (\ref{rate-8}) with the replacement $F_1
\to F_3$. (Note that this replacement has to be done in the definition
of $a$, despite the fact that there is no static effect from ${\cal
L}_3$.)

For the coupling to the gauge ghost condensate, we work in the regime
$g \gg m/E$ which, for relativistic neutrinos, holds for generic
values of $g$. For simplicity, we also assume $g \gg E/M$ since in the
opposite limit the theory coincides with ghost condensation. For the
dimension five operator of equation (\ref{gaugedint}) we find
\beq
\Gamma=\frac{g^{3}\sqrt{\beta}E}{3 \pi M^{2}}\,\mu^{2}, \qquad
-\frac{dE}{dt}= \frac{g^{4}\beta E^{2}}{4 \pi M^{2}}\,\mu^{2}, \qquad
\langle \cos\varphi\rangle=\frac{3}{5}.
\eeq
As mentioned in section
\ref{couplingneutrino}, these results are not mass suppressed as
they were in the ungauged case, so the dimension eight operators in the
gauged case should be subdominant. Yet, since the dimension five
operators can be forbidden by a symmetry, we quote the result for the
dimension eight operators as well:
\beq \label{rate-8-gauge}
\Gamma=\frac{g p^{3}}{15\pi\sqrt{\beta}M^2\,}a^2,\qquad
-\frac{dE}{dt}=\frac{g^{2}p^{4}}{12\pi M^2}\,a^2,\qquad
\langle \cos\varphi\rangle=\frac{1}{7}.
\eeq
Here, as for ghost condensation, ${\cal L}_2$ does not contribute and
for ${\cal L}_3$ the results can be obtained from (\ref{rate-8-gauge})
with the replacement $F_1\to F_3$.

We are now ready to use these results to constrain the sizes of
the Lorentz-violating operators.

\subsection{Bounds from SN1987A}
First, we will apply our results to neutrinos arriving at the Earth
from SN1987A.  Since the observed number of neutrinos is consistent
with existing supernovae models and we found that if a neutrino
radiates even a single Goldstone it will be thrown off its original
trajectory, we demand
\beq
\Gamma\;t_{87A}\lsim 1.
\eeq

Note that our results for ghost condensation can be trusted as long
as $E \ll M$. The observed neutrino spectrum from  SN1987A implies
that we need $M\gg 10\,\MeV$ yet the bounds from gravitational
experiments is $M\lsim 10 \;\MeV$ \cite{Arkani-Hamed:2003uy}. Thus,
our results in the case of ghost condensation should be taken as an
order of magnitude estimate subject to $O(1)$ corrections. In the
case of gauged ghost condensation, however, there is no difficulty
taking $M\gg 10\,\MeV$ \cite{Gauge-ghost}  so this issue does
not arise.

We adopt $M\sim 10\,\MeV$ and a generic neutrino mass of $m \sim
0.1\;\eV$ as reference values.  We fix the supernova neutrino energy
$E = 10 \, \MeV$.  With this choice we obtain bounds on the dimension
five operator in (\ref{rate-5}) 
\beq
\mu \lsim 10^{-11}\;{\GeV}\;\left(\frac{M}{10\;\MeV}\right)^{3/2}
\left(\frac{0.1\; \eV}{m}\right),
\eeq
and on the dimension eight operator in (\ref{rate-8}),
\begin{equation}
a \lsim 10^{-17}\left(\frac{M}{10\;\MeV}\right)^{3/2}.
\end{equation}
For couplings to the gauged ghost condensate, we use a reference value
$g\sim 10^{-3}$, and the bound from the dimension five operator gives
\beq
\mu\lsim 10^{-15}\;{\GeV}\left(\frac{M}{10\;\MeV}\right)
\left(\frac{10^{-3}}{g}\right)^{3/2}.
\eeq
Last, for the dimension eight coupling in gauged ghost condensation
we get
\beq
a \lsim 10^{-15} \left(\frac{M}{10\;\MeV}\right) \left(10^{-3} \over
g\right)^{1/2}. 
\eeq

We could also use the fact that the neutrinos from SN1987A arrived
with roughly their expected energy spectrum. That is, we could demand
that $\Delta E \lsim 10\;\MeV$, but this would not improve the bounds
already obtained. The reason is that $\Delta E / E =(T/E)(dE/dt)$ is
generically smaller than $\Gamma\, T$ by a factor of $E/M$ (in ghost
condensation) or $g$ (in gauged ghost condensation) which are both
assumed to be less than one in our analysis.

\subsection{Bounds from Cosmology}
\label{sec:CMBR}

Another way to probe Goldstone emissions is through cosmological
observables, in particular the CMBR.  In standard FRW cosmology,
neutrinos decouple from the baryon/photon plasma around the time of
nucleosynthesis.  When the acoustic oscillations of the CMBR are being
formed, the neutrinos free-stream relativistically from over- to
under-dense regions \cite{Bashinsky:2003tk}, and their energy density
scales like radiation as the universe expands \cite{CMBR-review}.
Non-standard interactions of neutrinos can affect the CMBR if they
either inhibit free-streaming or if they transfer energy to a sector
that redshifts differently than radiation.  Generally, any theory that
deviates from the standard picture can be tested with precision
measurements of the microwave background. The energy density in
neutrinos is roughly characterized by an effective number of neutrino
species $N_\nu$. Energy losses of neutrinos would appear as an
effective number of species that is different from three. (Similar
explorations are done in ``late-time'' neutrino models
\cite{Chacko:2003dt, Okui:2004xn}.)

In ghost condensation, the interaction between the Goldstone and the
neutrino is suppressed by a factor of $1/F$, so to a very good
approximation the neutrinos are free-streaming over the time scales
relevant for acoustic peak formation.  However, the effect of overall
energy transfer from neutrinos to Goldstones can be significant.

As we show in appendix~\ref{app_grav_eng}, in ghost condensation, the
gravitational energy stored in Goldstone bosons redshifts like cold
dark matter (CDM). Therefore, if a large number of Goldstones are created
from neutrinos, the energy density in the universe that redshifts like
radiation would be less than in the standard case.  Such a difference
would affect the CMBR observations made today.  Of course, a detailed
study of all cosmological observables is needed to fully understand
the effect of Goldstone radiation.  Below we only estimate bounds by
demanding that neutrinos do not lose too much energy to
Goldstones. That is, we demand that the total relative energy lost to
Goldstones is not very large:
\beq \label{CMBR-req}
r\equiv-\int
{1 \over E} {dE \over dt} \,dt \lsim 0.1.
\eeq

Note that the time interval we are interested here --- from neutrino
decoupling to
the formation of the CMBR --- is roughly equal to the travel time of
neutrinos from SN1987A to the earth.   In the previous section, we
demanded that not even a single Goldstone be emitted from SN1987A
neutrinos.
Here, we only demand that the emitted Goldstones do not drain too much
energy from cosmological neutrinos, so one might expect that the bound
from the CMBR would be weaker than from SN1987A.   However, it is still
interesting to study the CMBR, because if $M<10\;\MeV$ then all
neutrino energies from SN1987A would be outside of the ghost condensate
effective theory, and the bounds derived in the previous section could
not be trusted.  For the CMBR, though, the
energy of the neutrinos redshifts down to roughly an $\eV$
towards the formation of the CMBR, and therefore we
can still get some dynamic bounds for low values of $M$ as well.

In order to perform the integral in (\ref{CMBR-req}), we have to know
how the neutrino energy depends on time. Assuming a radiation
dominated universe, we have \cite{KolbandTurner}
\beq
t = C_1 T^{-2},
\qquad C_1 \approx 3 g_*^{-1/2} M_{\Pl} \sim \MeV^2~\sec \qquad
\Rightarrow \qquad dt=-2C_1 T^{-3} dT.
\eeq
The rate of energy loss can be written as
\beq
-{dE \over dt} = C_2 E^n ,
\eeq
where $C_2$ and $n$ depend on the specific operator we are considering.
  Taking an average neutrino energy of $E \sim 3T$ we get
\beq
\left. r\sim 2 \times 3^{n-1} C_1 C_2 \int T^{n-4}\,dT
\sim {2 \times 3^{n-1} C_1 C_2 \over n-3}
T^{n-3}\right|_{T_{max}}^{T_{min}},
\eeq
where the final step works
for $n \ne 3$ and the integration limits are roughly from the time of
neutrino decoupling $T_{max}=1\;\MeV$
to the formation of the CMBR  $T_{min}=1\;\eV$.

We can get a rough bound on the model parameters by using the
requirement
in (\ref{CMBR-req}). For the dimension five operator in 
(\ref{rate-5}), we have
\beq
C_2 = {m^2 \mu^2 \over 2 \pi M^4}, \qquad n=2, \qquad \mu \lsim
10^{-20}\;\GeV \left({M \over 10\;\eV}\right)^2  \left( \frac{0.1\;
\eV}{m} \right).
\eeq
For the dimension eight operator in (\ref{rate-8}),
\beq
C_2  = \frac{a^2}{6\pi M^4}, \qquad n = 6, \qquad a \lsim 10^{-10}
\left( \frac{M}{10 \; \MeV}  \right)^2.
\eeq

As we show in appendix~\ref{app_grav_eng}, the equation of state for
gauged ghost condensation is that of radiation. Therefore one may be
able to obtain only weak bounds in this case because the equation of
state for neutrinos and the longitudinal mode are identical to leading
order. On the level of the rough bounds with which we are concerned,
we cannot place additional bounds on gauged ghost condensation through
cosmological considerations.

\section{Discussion and Conclusions}

All experimental evidence to date confirms Lorentz invariance to be a
highly accurate symmetry of nature.  Of course, Lorentz invariance
could be violated, and the way it is violated would have definite
experimental implications.  In this paper, we have focused on the
possibility that general relativity correctly describes nature at all
energies below $M_{\Pl}$, but that both diffeomorphisms and Lorentz
invariance are spontaneously broken at some scale $M$.  However, it is
possible that general relativity is only one limit of the true theory
of gravity, in which case there are two additional possibilities.
First, Lorentz invariance could be an accidental symmetry of
elementary particle interactions that is simply absent at higher
energies.  Second, just as Galilean invariance is the small velocity
limit of Lorentz invariance, Lorentz invariance could be just be some
limit of a more fundamental symmetry.

Spontaneous Lorentz violation allows us to make definite experimental
predictions in the low energy effective theory below the scale $M$
without having to postulate a full theory of modified gravity in the
ultraviolet. In particular, by assuming that general relativity
holds, the existence of Lorentz-violating vevs and Goldstone bosons
are robust, and they lead to novel phenomena that can be used to probe
spontaneous Lorentz violation.

In this paper, we assumed that the Lorentz-violating sector coupled
only to neutrinos, and we studied how neutrinos can place bounds on
the specific models of ghost condensation and gauged ghost
condensation.  Static effects arose from the presence of a
Lorentz-violating vev.  The dominant static effect is a modification
of the neutrino dispersion relation, which has been extensively
studied in the literature.  Here, we focused on dynamic effects from
spontaneous Lorentz violation.  Like any spontaneously broken
symmetry, spontaneously broken Lorentz symmetry is associated with a
massless Goldstone boson.  Every static interaction is accompanied by
a novel neutrino-Goldstone coupling.  We have seen that when neutrinos
travel in vacuum they can lose energy and change their direction of
motion due to Goldstone emission.  This Goldstone-\v{C}erenkov effect
can be used to probe the mechanism of spontaneous Lorentz violation.

We identified two kinds of observables that can be used to probe this
dynamic effect.  First, we looked at data on neutrinos from
astrophysical sources, in particular, from SN1987A.  If neutrinos from
SN1987A emitted Goldstone bosons on their way to the Earth, they would
have deflected away and the number of neutrinos to arrive at the Earth
would have changed. Because the data tells us that the neutrinos
travel from the supernova basically without interaction, we can put
stringent constraints on the size of Lorentz violation. We also
applied cosmological considerations to put bounds on the model
parameters. We know that the CMBR power spectrum agrees with the
presence of three standard neutrino species. Significant energy
transfer from neutrinos into the Goldstone field would affect standard
cosmology, and would therefore be inconsistent with observations.

\begin{table}[t]
\caption{Summary of bounds on Lorentz-violating
couplings. In all cases only rough estimates are given. The first
group corresponds to bounds from static effects on non-universal
(first two entries) and universal (third and fourth entries)
Lorentz-violating couplings. In the second group bounds on the
universal coupling are given both for ghost condensation (fifth and
sixth entries) and gauged ghost condensation (last entry). Here, $M$
is the scale of spontaneous Lorentz violation, $m$ is the neutrino
mass, and $g$ is the gauge coupling in gauged ghost condensation. N/A
is given when no bounds can be obtained or when bounds are unavailable
but are expected not to be significant. See the text for more details.
\label{tab:bounds}}
\vspace*{3mm}
\begin{tabular}{|c||c|c|} \hline\hline
& ~~$\mu$ [$\GeV$]~~ & ~~~$a$ [number] ~~~
\\[2pt] \hline
Static bounds & & \\ \hline
~Atmospheric (non-universal) ~~~ & $10^{-22}$ & $10^{-24}$\\
~Solar/Kamland (non-universal)~ & $10^{-20}$ & $10^{-18}$  \\
SN1987A &  N/A & $10^{-8}$ \\
~Decay kinematics~ & $10^{-6}$ & N/A \\[2pt] \hline
Dynamic bounds & & \\ \hline
~SN1987A (ghost condensation)~&
~$10^{-11}\left[{M \over 10\;\MeV}\right]^{3/2}
\left[{0.1\;\eV \over m}\right]$~&
~$10^{-17}\left[\frac{M}{10\; \MeV}\right]^{3/2}~$ \\[5pt]
CMBR (ghost condensation)~ & ~$10^{-22}\left[{M \over
1\;\eV}\right]^2\left[{0.1\;\eV \over m}\right]$~ &
  ~$10^{-11}\left[{M \over 1\;\MeV}\right]^2$~  \\[5pt]
~SN1987A (gauged ghost condensation)~&
~$10^{-15}\left[\frac{M}{10\;\MeV}
\right]\left[\frac{10^{-3}}{g}\right]^{3/2}$~ & 
~$10^{-15} \left[\frac{M}{10\;\MeV}\right] \left[10^{-3} \over
g\right]^{1/2}$~ 
   \\[5pt]
\hline\hline
\end{tabular}
\end{table}

We summarize the bounds on the coupling of neutrinos to the
Lorentz-violating sector in Table~\ref{tab:bounds}. We quote bounds on
two kind of operators, the dimension five operator in (\ref{lowest})
and the dimension eight operator in (\ref{dim-eight}). We quote bounds
from static and dynamic effects. The main conclusion of our paper can
be read off the table. We see that dynamic effects can be much more
effective in probing Lorentz violation than static effects, especially
for flavor-universal couplings. That is, ignoring the presence of the
Goldstone boson is generically not justified.

Future experiments can be used to further probe Lorentz violation,
enabling us to better constrain the degree of Lorentz violation or to
discover it. Detecting neutrinos from cosmological distances, as in
gamma ray bursts, will provide much stronger probes as such neutrinos
travel very long distances.  A close by supernova observed with a high
statistics neutrino signal would also be very useful to refine our
rough bounds. Finally, the study of the cosmological implication of
the neutrino-Goldstone interaction has to be refined. Here, we gave a
very rough estimate of the effect using only energy-loss
arguments. Clearly, a detailed study will be useful to fully
understand how such effects can be discovered or further bounded.

\acknowledgments

We thank N.~Arkani-Hamed, H.-C.~Cheng, M.~Cirelli, A.~Cohen,
P.~Creminelli, H. George, S.~Glashow, M.~Luty, S.~Mukohyama,
A.~E.~Nelson, T.~Okui, V.~Sanz, Y.~Shadmi and M.~Zaldarriaga, for
helpful discussions.

\appendix

\section{Gauged Ghost Condensation}
\label{sec:ggc}

Here we give some details of gauged ghost condensation, following
\cite{Gauge-ghost}. Starting with the Lagrangian in equation
(\ref{ggclagrange}), we derive the dispersion relation and
polarization of the Goldstone mode in gauged ghost condensation.  For
ease of discussion, we work out the details in the limit $k \ll gM$,
and then quote results in the general case to see how ghost
condensation is recovered from gauged ghost condensation in the $g
\rightarrow 0$ limit.  The gauged ghost Lagrangian is:
\beq  \label{ggclagrange2}
\mathcal{L} =
-\frac{1}{4g^2}F_{\mu\nu}^2 + \frac{1}{8M^4}\left((D_\mu \phi)^2 -
M^4 \right)^2 - \frac{\beta}{2M^2}(\partial^\mu D_\mu \phi)^2 -
\frac{\beta'}{2M^2}(\partial_\mu D_\nu \phi)^2  + \ldots.
\eeq
First, we can go to a ``unitary'' gauge where $\phi \equiv M^2 t$.
Keeping only terms quadratic in $A_\mu$ we have
\beq \label{ggcunitary}
\mathcal{L} = -\frac{1}{4g^2}F_{\mu\nu}^2 + \frac{M^2}{2}A_0^2 -
\frac{\beta}{2}(\partial^\mu A_\mu)^2 -
\frac{\beta'}{2}(\partial_\mu A_\nu)^2  + \ldots,
\eeq
We can absorb $\beta'$ into a redefinition of $g$ and $\beta$, so we
set $\beta' = 0$. Since the mode $A_0$ has mass $gM$ we can integrate it out.  Ignoring
$k/M$ corrections to the dispersion relations of the massless
polarizations, we can simply set $A_0 = 0$, and the $A_i$ Lagrangian
is
\beq \mathcal{L} = \frac{1}{2g^2} (\partial_0 A_i)^2 -
\frac{1}{2g^2}(\partial_i A_j)^2 + \frac{1}{2g^2}\left(1 - \beta g^2
\right)(\partial^i A_i)^2.
\eeq
Expanding $A_i$ in plane waves with $k_\mu = (E,0,0,k)$, the classical
polarization vectors are as follows. The transverse modes have Lorentz
invariant dispersion relations and are
\beq
E^2 = k^2; \qquad    \epsilon_i^1 = (g,0,0),
\qquad  \epsilon_i^2 = (0,g,0).
\eeq
The longitudinal mode, however, has a Lorentz-violating
dispersion relation
\beq \label{intoutpolar}
E^2 = \beta g^2 k^2 ; \qquad  \epsilon_i^L = (0,0,g) = g k^i/|k|.
\eeq
To verify the normalization of the polarizations, we calculate the
vector field propagator
\beq
\langle A_i A_j \rangle = \frac{g^2}{E^2 - k^2} \left(\delta_{ij} -
(1-\beta g^2) \frac{k_i k_j}{E^2 - \beta g^2 k^2}\right).
\eeq
As expected from the cutting rules, we find
\beq
\left. \langle A_i A_j \rangle (E^2 - k^2) \right|_{E = k} =
\sum_{n=1,2} \epsilon_i^n \epsilon_j^{n*}, \qquad \left.
\langle A_i A_j \rangle
(E^2 - \beta g^2 k^2) \right|_{E = \sqrt{\beta} g k} =
\epsilon_i^L \epsilon_j^{L*}.
\eeq

While not directly relevant to our discussion, it is instructive to
consider the more general case with generic values of $gM$ and
$k$. This will allow us to see how original ghost condensation is
recovered in the $g \rightarrow 0$ limit. In that case, we need to
work with the full $A_\mu$ field and cannot integrate out $A_0$.
Using equation (\ref{ggcunitary}) with $\beta' = 0$, there are still
two transverse polarization vectors with relativistic dispersion
relations:
\beq
E^2 = k^2; \qquad    \epsilon_\mu^1 = (0,g,0,0),
\quad  \epsilon_\mu^2 = (0,0,g,0).
\eeq
There are also two ``longitudinal'' modes.  The first is a ghost
excitation (\emph{i.e.}\ the residue at the relevant pole of the
propagator is negative), but it has a massive dispersion relation
\beq
E_{\rm ghost}^2 = \frac{M^2}{\beta} + (2-g^2 \beta) k^2 +
\mathcal{O}(k^4/M^2),
\eeq
so for $E \ll M/\sqrt{\beta}$, this unhealthy mode is never excited.
The Goldstone longitudinal mode has dispersion relation
\beq
E_L^2 = \beta g^2 k^2 + \beta (1-g^2 \beta)^2 \frac{k^4}{M^2} +
\mathcal{O}(k^6/M^4),
\eeq
which matches the ghost condensate dispersion relation in the $g
\rightarrow 0$ limit.  The normalized polarization vector is
\beq
\epsilon_\mu^L = \left( \frac{E}{g M^2}  \left((1-g^2 \beta)k +
\mathcal{O}(k^2/M^2)   \right),0,0,g + \mathcal{O}(k^2/M^2) \right),
\eeq
which indeed reproduces equation (\ref{intoutpolar}) up to
$\mathcal{O}(k^2/M^2)$ corrections.  However, we see that the zero
gauge coupling limit is singular if we do an expansion in $k/M$ and
then try to take $g \rightarrow 0$.  This is to be expected, because
with $g = 0$ the Goldstone energy scales as $k^2$, but for finite $g$,
$E \sim k$.  Expanding in $g$ first, and then considering $(k/M)$
corrections we get
\beq
\epsilon_\mu^L = \left( \frac{E}{M}
\left(1 + \mathcal{O}(k^2/M^2) \right) +
\mathcal{O}(g^2),0,0, \frac{k}{M} +
\mathcal{O}(k^2/M^2)+ \mathcal{O}(g^2) \right).
\eeq
And we see that at the level of longitudinal polarizations
\beq
D_\mu \phi \equiv M^2 \delta_\mu^0 +
 M A^L_\mu \quad \Rightarrow \quad \partial_\mu \phi =
M^2 \delta_\mu^0 + \partial_\mu \pi,
\eeq
in the $g \rightarrow 0$ limit.


\section{Goldstone-\v{C}erenkov Calculation}
\label{sec:calcappen}

In this appendix, we give details of our dynamic calculations. We
consider an ultra-relativistic (in the ether rest frame) neutrino with
mass $m$ that emits a Goldstone
\beq
\nu(p) \to \nu(q) + \pi(k),
\eeq
as in figure \ref{decayfig}.  We consider the case where the initial
and final neutrino is the same. We use
\beq
p_{\mu}=(E_{in},0,0,p), \qquad
q_{\mu}=(E_{out},q \sin\varphi,0,q \cos\varphi),\qquad
k_{\mu}=(E_{\pi},k \sin\theta,0,k \cos\theta).
\eeq
We define for a general four vector $V_{\mu}$,
$\tilde{V}_{\mu}=(V,-\vec{V})$.  In the following, we work to leading
order in $m/E$. Then, $E_{in}=p$ and $E_{out}=q$.  We express all
constrained kinematic variables in terms of $p$ and $\theta$.

The differential rate is given by the standard formula \cite{peskin}
\beq
d \Gamma = {d^3k \over (2\pi)^3} {d^3q \over (2\pi)^3}
 {1 \over (2 E_{in})(2E_{out})(2E_{\pi})}  \left|{\cal M}\right|^2
(2\pi)^4 \delta^{(4)}(p-q-k).
\eeq
We always work in regions where the effective theory is valid. Then,
$E_\pi \ll k$ and to leading order in $E_\pi/k$ we have $p=q$ [see
(\ref{pandE-global}) and (\ref{pandE-gauge})]. It turns out to be more
convenient to use the $\delta^{(3)}$ function to eliminate
$\vec{q}$. Performing the trivial integration over the azimuthal angle
we have an expression that depends only on $p$, $k$ and $\cos\theta$
\beq \label{master-on}
d \Gamma = {k^2 \,dk\, d(\cos\theta)
 \over 16 \pi \,p^2\,E_{\pi}(k)}  \left|{\cal M}\right|^2
{ \delta(k-\kon) \over  f'(\kon)}.
\eeq
Here,
\beq
f(k)=E_\pi(k)+E_{out}(k)-E_{in}(k),
\eeq
is the energy conservation condition where the solution of the
on-shell condition is denoted by $\kon$.  We also used the fact that
when $f(x)=0$ has a single solution $x=x_0$ then
\beq
\int ... \,\delta(f(x))dx = \int ... \,{\delta(x-x_0) \over
|f'(x_0)|} dx.
\eeq

In order to calculate the rate we need to know $f(k)$ (that is, the
dispersion relation of the Goldstone) and the matrix element, ${\cal M}$.
Both are model dependent, and below we continue the calculations in
the different models we are considering.


\subsection{Ghost Condensation}
We use the Goldstone dispersion relation from (\ref{pion-dis}). We
work in the limit where the effective theory is valid, that is, when
$k/M\ll 1$. While we calculate to subleading order in $k/M$, we report
only the leading order results.

The energy and momentum conservation equations for ghost condensation
are
\beq \label{pandE-global}
p=q+\sqrt{\beta} {k^2 \over M}, \qquad
q^2=p^2+k^2 - 2 p k \cos\theta.
\eeq
The solutions to first order in $k/M$ are
\beq \label{on-shell-global}
q=p,\qquad \kon = 2 p \cos\theta, \qquad
\cos\varphi=-\cos 2\theta, \qquad f'(\kon)=\cos\theta,
\eeq
and the allowed values for $\cos\theta$ are $0 \le \cos\theta\le 1$.
We then get from (\ref{master-on})
\beq \label{master-on-global}
d \Gamma = {M \over 16 \pi p^2 \,\sqrt{\beta} \cos\theta}
\left|{\cal M}\right|^2  d(\cos\theta).
\eeq

To continue further, we need to specify the interaction. We start with
the dimension five operator in (\ref{lowest-int}). This operator leads
to a $\nu-\nu-\pi$ vertex that is proportional to
$\slashed{k}/F$. Thus, for the amplitude we get
\beq
{\cal M} = \frac{1}{F}\bar{u}(q)\slashed{k}P_{L}u(p)=
\frac{m}{F}\bar{u}_{q}\gamma^{5} u_{p},
\eeq
where $P_{L}=(1-\gamma^{5})/2$ and the equation of motion of the
fermion was used in the last step. (While static effects modify the
neutrino dispersion relation, this change is a higher order effect and
we verify that it is indeed negligible when considering the neutrino
kinematics.)  Note that the amplitude is proportional to the neutrino
mass. This can be understood by the fact that for a massless neutrino,
the dimension five interaction is not physical because it can be
rotated away by a field redefinition. Using standard spin sum and
trace technology we get
\beq \label{ms-global}
\sum |\mathcal{M}|^{2}=\frac{m^{2}}{F^{2}}
\Tr\left(\slashed{p}\gamma^{5}\slashed{q}\gamma^{5}\right)=
-\frac{4m^{2}}{F^{2}}p_{\mu} q^{\mu}=\frac{2m^{2}k^{2}}{F^{2}}
=\frac{8m^{2}p^{2}\cos^2\theta}{F^{2}},
\eeq
where in the last step we use the on-shell condition,
(\ref{on-shell-global}).  While we do not present here the polarized
results explicitly, we note that the interaction requires a spin
flip.

Using (\ref{ms-global}) in (\ref{master-on-global}) we can calculate
the observables we are after. In particular, we calculate the total
width, $\Gamma$, the average energy lost to the Goldstone, $-dE/dt$,
and the average deflection of the neutrino, $\langle
\cos\varphi\rangle$. To leading order in $p/M$ we get
\beqa
\Gamma&=&\int d\Gamma=\frac{Mm^{2}}{2\pi\sqrt{\beta} F^{2}}
\int_0^1 \cos\theta\, d(\cos\theta)
=\frac{Mm^{2}}{4\pi\sqrt{\beta} F^{2}},\\
-\frac{dE}{dt}&=&\int d\Gamma E_{k}=\frac{m^{2}p^{2}}{2\pi
F^{2}},\nonumber \\
\langle \cos\varphi\rangle&=&\frac{\int d\Gamma \cos\varphi}{\int
d\Gamma}
=\frac{4\sqrt{\beta}p}{3M}\sim 0. \nonumber
\eeqa

Next we repeat the calculation using the dimension eight operators. In
that case, we set the neutrino mass to zero, since, unlike the dimension
five case, it is not required in order to have a nonzero effect.  We start
with $\mathcal{L}_{1}$ in (\ref{dim-eight-int}), where the
$\nu-\nu-\pi$ vertex is
\beq
\frac{M^{2}}{F_{1}^{4}}\left(\gamma^{0}k_{\mu}p^{\mu}+
\slashed{k}p^{0}-2k^{0}\slashed{p}\right)P_{L}.
\eeq
The last two terms vanish for on-shell neutrinos and thus
the amplitude is
\beq
{\cal M}=\frac{M^{2}}{F_{1}^{4}}\bar{u}(q)\gamma^{0}k_{\mu}p^{\mu}
P_{L}u(p).
\eeq
Note that this amplitude conserves spin.  Summing over spins we get
\beq \label{ms-eight-global}
\sum|\mathcal{M}|^{2}=\frac{M^{4}}{F_{1}^{8}}
\left(k_{\mu}p^{\mu}\right)^{2}
\Tr\left(\slashed{q}\gamma^{0}P_{L}\slashed{p}\gamma^{0}\right)
=\frac{2M^{4}}{F_{1}^{8}}\left(k_{\mu}p^{\mu}\right)^{2}
\left(p_{\nu}\tilde{q}^{\nu}\right)
=\frac{16M^{4}}{F_{1}^{8}}p^{6}\cos^{4}\theta\, \sin^{2}\theta,
\eeq
where in the last step we used the on-shell condition.  Using
(\ref{ms-eight-global}) in (\ref{master-on-global}) we get to leading
order in $p/M$
\beq \label{dim-8-res}
\Gamma=\frac{M^{5}p^{4}}{12\pi\sqrt{\beta} F_{1}^{8}}, \qquad
-\frac{dE}{dt}=\frac{M^{4}p^{6}}{6\pi F_{1}^{8}},\qquad
\langle \cos\varphi\rangle=\frac{8\sqrt{\beta}p}{5M}\sim 0.
\eeq

The calculation for the other dimension eight operators is very
similar. In the massless limit $\mathcal{L}_{2}$ in (\ref{dim-eight})
gives no effects for on-shell neutrinos because it is proportional to
the trace of the energy momentum tensor.  For $\mathcal{L}_{3}$, we go
through essentially the same analysis as for $\mathcal{L}_{1}$ and
find
\begin{equation}
|\mathcal{M}|^{2}=\frac{16M^{4}}{F_{3}^{8}}p^{6}\cos^{4}\theta\,
\sin^{2}\theta,
\end{equation}
which leads to the same result as in (\ref{dim-8-res}) with the
replacement $F_1 \to F_3$. Note that there is no fundamental reason
that ${\cal L}_1$ and ${\cal L}_3$ give the same results.


\subsection{Gauged Ghost Condensation}
We move to perform the calculation in the gauged case. The process we
are after is the same as for the global case but in the gauged case
the emitted boson is a vector. The neutrino can emit the longitudinal
mode of the vector since that is the one that has a Lorentz-violating
dispersion relation. The dispersion relation of the longitudinal mode
is given in equation (\ref{gauge-disp}). We assume
\beq
g\gg {k \over M}, \qquad g\sqrt{\beta}\gg {m\over p},
\qquad  g\sqrt{\beta}\ll 1,
\eeq
and keep only leading contributions in these small quantities.

The energy and momentum conservations for the gauged case are
\beq  \label{pandE-gauge}
p=q+\sqrt{\beta} g k, \qquad
q^2=p^2+k^2 - 2 p k \cos\theta.
\eeq
The solutions to first order in the small quantities is the same as
for ghost condensation (\ref{master-on-global}) (differences arises in
high order terms).  Going through the same steps as in the previous
subsection, we derive from (\ref{master-on})
\beq \label{master-on-gauged}
d\Gamma=\frac{1}{8\pi\,p\, g \,\sqrt{\beta}} \left|{\cal M}\right|^2
d\left(\cos\theta\right).
\eeq

The vertex can be  derived from (\ref{gaugedint}) by
substituting the correctly normalized polarization vector for the
longitudinal mode given in (\ref{long-polarization}) and it is
proportional to $\slashed{\epsilon}M/F$. For the
amplitude we then get
\begin{equation}
{\cal M} = \frac{M}{F}\bar{u}(q)\slashed{\epsilon} P_{L}u(p)
= \frac{M}{F}g\bar{u}(q)\hat{k}\cdot\vec{\gamma}P_{L}u(p)
=-\frac{g^{2}\sqrt{\beta}M}{F}\bar{u}(q)\gamma^{0}P_{L}u(p),
\end{equation}
where in the last step we use the fact that in the massless limit
$\bar{u}(q)\slashed{k}P_{L}u(p)=0$. Note that unlike ghost
condensation the rate is not proportional to the neutrino mass. This
is to be expected as in the gauged case the dimension five interaction
cannot be rotated away in the massless limit.
Summing over spins we get
\beq \label{spin-sum-gg}
|\mathcal{M}|^{2}=\frac{g^{4}\beta M^{2}}{F^{2}}
\Tr\left(\slashed{q}\gamma^{0}P_{L}\slashed{p}\gamma^{0}\right)
=\frac{2g^{4}\beta M^{2}}{F^{2}}\left(p_{\mu}\tilde{q}^{\mu}\right)
=\frac{4g^{4}\beta M^{2}}{F^{2}}p^{2}\sin^{2}\theta,
\eeq
where we used the on-shell condition in the last step.  Using
(\ref{spin-sum-gg}) in (\ref{master-on-gauged}) we can calculate the
relevant observables and we get
\beq
\Gamma=\frac{g^{3}\sqrt{\beta}M^{2}p}{3\pi F^{2}},\qquad
-\frac{dE}{dt}=\frac{g^{4}\beta M^{2}p^{2}}{4\pi F^{2}},\qquad
\langle \cos\varphi\rangle=\frac{3}{5}.
\eeq

Next we consider the dimension eight operators in the gauged case.
The vertex we get from covariantizing the operator in equation
(\ref{dim-eight}) similar to (\ref{gaugedint}) and using the
polarization vector for the longitudinal mode we find the amplitude
\begin{equation}
{\cal M} =\frac{gM^{3}}{F_{1}^{4}}
\bar{u}(q)\left(-\gamma^{0}\vec{p}\cdot\hat{k}+p^{0}\hat{k}
\cdot\vec{\gamma}\right)P_{L}u(p)=
-\frac{M^{3}g}{F_{1}^{4}k}p_{\mu}\tilde{k}^{\mu}\bar{u}(q)\gamma^{0}P_{L}u(p),
\end{equation}
where in the last step we trade $\vec{k}\cdot\vec{\gamma}$ for
$-k^{0}\gamma^{0}$.  Using a spin sum we get
\beq
|\mathcal{M}|^{2}=\frac{M^{6}g^{2}}{k^{2}F_{1}^{8}}
\left(p_{\mu}\tilde{k}^{\mu}\right)^{2}
\Tr\left(\slashed{q}\gamma^{0}P_{L}\slashed{p}\gamma^{0}\right)
=\frac{2M^{6}g^{2}}{k^{2}F_{1}^{8}}
\left(p_{\mu}\tilde{k}^{\mu}\right)^{2}\left(p_{\nu}\tilde{q}^{\nu}\right)
=\frac{4M^{6}g^{2}}{F_{1}^{8}}p^{4}\sin^{2}\theta \cos^{2}\theta,
\eeq
where we used the on-shell condition in the last step.
Using the above in (\ref{master-on-gauged}) we find
\beq
\Gamma=\frac{M^{6}g p^{3}}{15\pi\sqrt{\beta}F_{1}^{8}},\qquad
-\frac{dE}{dt}=\frac{M^{6}g^{2}p^{4}}{12\pi F_{1}^{8}},\qquad
\langle \cos\varphi\rangle=\frac{1}{7}.
\eeq

Finally, we perform the calculation for the remaining dimension eight
operators. As in ghost condensation, $\mathcal{L}_{2}$ does not
contribute for on-shell neutrinos.  For $\mathcal{L}_{3}$ we found
that the results are the same as for $\mathcal{L}_{1}$ up to the
replacement $F_1 \to F_3$.


\section{Particle Physics vs. Gravitational Energy}
\label{app_grav_eng}

In section \ref{sec:CMBR}, we used an energy-loss argument to bound
the amount of energy transfered from neutrinos to Goldstone bosons
prior to the formation of the CMBR. The key to this argument is that
neutrinos and Goldstone bosons redshift differently. In particular, in
the appropriate approximation, neutrinos redshift like radiation and
Goldstone bosons like CDM. In this appendix we verify this, while
pointing out an interesting subtlety in calculating the equation of
state for the Goldstone boson.

The usual way \cite{KolbandTurner} to figure out the equation of state
for a particle is to note that in a FRW background, any
energy-momentum tensor has the form of a perfect fluid.  Thus,
conservation of stress-energy provides the first law of thermodynamics
\beq
d(\rho\, a^3) = - \mathcal{P}\, d(a^3),
\eeq
where $\rho$ is the density and $\mathcal{P}$ the pressure. This law
yields the physical intuition that energy in a comoving volume is
equal to minus the pressure times the same volume.  Assuming the
simple equation of state $\mathcal{P} = w \rho$, the first law
thermodynamics gives the evolution of the energy density as
\beq
\rho \sim a^{-3(1 + w)}.
\eeq
In general, for a normal species particle whose dispersion relation goes as
$E \sim k^n$, the corresponding to the equation of state is
\beq
w = \frac{n}{3}.
\eeq
For cold dark matter where there is no energy redshift, $w=0$, and for
relativistic neutrinos where $E \sim k$, $w = 1/3$.  One might then
conclude that for the Goldstone boson where $E \sim k^2$, the equation
of state would be $w = 2/3$.

However, this argument assumes that the equivalence principle is
observed.  In ghost condensation, however, time diffeomorphisms are
spontaneously broken, so there is no reason to expect that the current
associated with space-time translations (the ``particle physics''
energy) would be the same as the stress-energy tensor (the
gravitational energy) \cite{Arkani-Hamed:2003uy}.  To leading order in
$k/M$, the stress energy tensor for the ghost condensate is
\beq
\label{realtmunu}
T_{\mu\nu} = \frac{1}{2M^4} \left( (\partial_\mu \phi)^2 - M^4  \right)
\partial_\mu \phi \partial_\nu \phi - \frac{2\beta}
{M^{2}}\partial_{\mu}\partial_{\nu}\phi\left(\partial^{\alpha}
\partial_{\alpha}\phi\right)-g_{\mu\nu}\mathcal{L},
\eeq
and expanding around the vacuum, $\partial_\mu \phi = M^2 \delta_\mu^0 +
\partial_\mu \pi$, the gravitational energy density for Goldstone is
\beq
\rho_{\rm grav}=T_{00} = M^2 \dot{\pi} + 2 \dot{\pi}^2 -
\frac{1}{2}(\nabla \pi)^2 +
\frac{\beta}{2M^{2}}\left(\nabla^{2}\pi\right)^{2} + \ldots
\eeq
which looks nothing like the particle physics energy density for the
Goldstone:
\beq
\rho_{\rm particle} = \frac{1}{2}\dot{\pi}^2 +
\frac{\beta}{2M^2}(\nabla^2 \pi)^2. 
\eeq
As for the pressure of the Goldstone, we find
\begin{equation}
3\mathcal{P}=\sum_{i}
T_{ii}=\frac{3}{2}\dot{\pi}^{2}-\frac{3\beta}{2M^{2}}\left(\partial_{\alpha}\partial^{\alpha}\pi\right)^{2}+\frac{2\beta}{M^{2}}\left(\nabla^{2}\pi\partial_{\alpha}\partial^{\alpha}\pi\right)+\ldots.
\end{equation}
Now as we impose the Goldstone equation of motion and average over time, the $\dot{\pi}$ term in equation
(\ref{realtmunu}) averages to zero and several terms cancel each other, leaving
\begin{equation}
\rho_{\rm grav}=-\frac{1}{2}\left(\nabla\pi\right)^{2}+\ldots\qquad \mathcal{P}=-\frac{2\beta}{3M^{2}}\left(\nabla^{2}\pi\right)^{2}+\ldots
\end{equation}
which leads to the equation of state
\begin{equation} \label{w-Goldstone}
w_{\rm Goldstone}=\frac{4\beta}{3}\frac{k^{2}}{M^{2}}\sim 0.
\end{equation}

There is another, more intuitive way to figure out the equation of
state for the Goldstone boson using equation (\ref{realtmunu}). A
relativistically normalized plane wave of $\pi$ in a box of volume $V$
takes the form
\beq
\pi(x,t) = \frac{1}{\sqrt{E V}} e^{i E t - i \vec{k}\cdot \vec{x}},
\eeq
where $E = \sqrt{\beta} k^2 / M$. Plugging into $T_{00}$ and averaging
over some reasonable amount of time we have
\beq
T_{00} = \frac{1}{E V} \left(\frac{5}{2}E^2 - \frac{1}{2} k^2 \right)
= - \frac{M}{2\sqrt{\beta}V} + \mathcal{O}(k/M).
\eeq
We see that $T_{00}$ for a Goldstone plane wave is the same as for a
non-relativistic particle plane wave with negative gravitational mass
$-M/\sqrt{\beta}$.\footnote{Since we are dealing with a theory with
broken time diffeomorphisms, one should not be concerned by the
presence of an anti-gravitating field. For the rough estimates
presented here, all we care about is that the gravitational energy
carried by Goldstones differs from neutrinos. A complete analysis of
CMBR signals of Lorentz violation would require a full understanding
of the gravitational dynamics of the Goldstone.}

As the universe expands, the only redshift in $T_{00}$ comes from the
volume factor, so we recover (\ref{w-Goldstone}).  Thus performing the
analysis both ways leads us to the same conclusion, namely that indeed
the equations of state of neutrino and the Goldstone boson
differ. Despite having an $E \sim k^2$ dispersion relation, the
gravitational energy in Goldstone radiation redshifts like cold dark
matter with negative mass.

We can go through a similar analysis in the case of gauged ghost
condensation. In that case, we expect the energy carried by the
longitudinal mode to scale like radiation. The reason is that in
gauged ghost condensation the gravitational and particle energies
are the same up to small corrections \cite{Gauge-ghost}. Below we show
that this is indeed the case.

The energy momentum tensor for the gauged case is given by
\begin{equation}
T^{\mu\nu}=-\frac{1}{g^{2}}F^{\mu\alpha}F^{\nu}_{\alpha}+
\frac{1}{2M^{4}}D^{\mu}\phi D^{\nu}\phi\left(\left(D\phi\right)^{2}-
M^{4}\right)-\frac{2\beta}{M^{2}}
\partial^{\mu}D^{\nu}\phi\left(\partial^{\alpha}D_{\alpha}\phi\right)-
g^{\mu\nu}\mathcal{L}.
\end{equation}
Now we go to the unitary gauge where $D_{\mu}\phi=\delta_{\mu
0}M^{2}+MA_{\mu}$ and we work in the $g\gg k/M$ limit where the
$A_{0}$ mode is heavy and can be integrated out by setting it to its
equation of motion. To leading order this has the effect of setting
the contribution from the $\left(\left(D\phi\right)^{2}-M^{4}\right)$
to zero so we find
\begin{equation} \label{Tzeroghost}
T^{00}=\frac{1}{2g^{2}}\left(\sum_{i}\left(F^{0i}\right)^{2}+\sum_{i>j}\left(F^{ij}\right)^{2}\right)+\frac{\beta}{2}\left(\vec{\nabla}\cdot\vec{A}\right)^{2}+\ldots
\end{equation}
where $A_{0}$ has been set to zero to leading order by its equation of
motion. We also find
\begin{equation}
\sum_{i}T^{ii}=T^{00}
\end{equation}
so that
\begin{equation} \label{ggeos}
w_{\rm gauged\; ghost}=\frac{1}{3},
\end{equation}
and the gauged ghost redshifts like radiation. Note that the result we
have found includes the transverse modes also, which one would
intuitively expect to redshift like radiation. The main point is that
all modes --- in particular the longitudinal mode which has a
Lorentz-violating dispersion relation --- have the equation of state
characteristic of radiation.

In fact, we can show this another way, by plugging the 
polarization vector for the longitudinal mode into (\ref{Tzeroghost})
to isolate its contribution to the gravitational energy density.
The relativistically normalized plane wave is
\begin{equation}
\vec{A}_{L}=\frac{1}{\sqrt{EV}}\,\vec{\epsilon}_{L}\,e^{i E t - i
\vec{k}\cdot \vec{x}}.
\end{equation}
We find to leading order that
\begin{equation}
\rho_{\rm grav}=\frac{g^{2}k^{2}\beta}{EV}=\frac{gk\sqrt{\beta}}{V},
\end{equation}
so the longitudinal mode redshifts like radiation and has the equation
of state of (\ref{ggeos}).


\end{document}